\pdfoutput=1

\documentclass[11pt,twoside,a4paper,cmspaper,final,collab]{cms-tdr}
\def\svnVersion{68535:68865}\def\cmsCernNo{2011-089}\def\cmsCernDate{\today}\def\cmsMessage{Submitted to Physical Review Letters}

\begin{document}\cmsNoteHeader{EXO-11-001}

\hyphenation{had-ron-i-za-tion}
\hyphenation{cal-or-i-me-ter}
\hyphenation{de-vices}
\RCS$Revision: 68865 $
\RCS$HeadURL: svn+ssh://alverson@svn.cern.ch/reps/tdr2/papers/EXO-11-001/trunk/EXO-11-001.tex $
\RCS$Id: EXO-11-001.tex 68865 2011-07-15 14:59:30Z alverson $
\cmsNoteHeader{EXO-11-001} % This is over-written in the CMS environment: useful as preprint no. for export versions
\title{\texorpdfstring{Search for Three-Jet Resonances in \Pp\Pp\ Collisions\\ at $\sqrt{s}$ = 7 TeV}{Search for Three-Jet Resonances in pp Collisions at sqrt(s) = 7 TeV}}

\date{\today}

\abstract{
    A model-independent search for three-jet hadronic resonance production in \Pp\Pp\ collisions at
    a center-of-mass energy
    of 7 TeV has been conducted by the CMS Collaboration at the LHC, using
    a data sample corresponding to an integrated luminosity of 35~pb$^{-1}$. Events
    with high jet multiplicity and a large scalar sum of jet transverse momenta are
    analyzed. The number of expected standard model background events is found to
    be in good agreement with the observed events. Limits are set on a model
    describing the production of $R$-parity-violating supersymmetric gluino pairs,
    and gluino masses in the range of 200 to 280 GeV/$c^2$ are excluded at a 95\%
    confidence level for the first time.
}

\hypersetup{%
pdfauthor={CMS Collaboration},%
pdftitle={Search for Three-Jet Resonances in pp Collisions at sqrt(s) = 7 TeV},%
pdfsubject={CMS},%
pdfkeywords={CMS, physics, software, computing}}

\maketitle %maketitle comes after all the front information has been supplied

Studies of multijet final states, although experimentally challenging at
hadron colliders, are sensitive to many extensions of the standard model (SM).
For example, variations of technicolor models, resulting in heavy colored fermions
that transform as octets under $SU(3)_\textrm{c}$, have been proposed in a variety
of forms ~\cite{tcgut,exoquarks,unichiral,altchiral}. Other models
incorporate $R$-parity-violating (RPV) decays of supersymmetric gluinos to
three-quark final states, where the gluino represents a colored adjoint
Majorana fermion~\cite{sixjets,multijets,reuven}.
In all cases, these high mass resonances can be pair-produced,
yielding a six-jet final state. Recent results from the Tevatron provide
limits on gluino RPV decays for masses below 144~GeV/$c^2$~\cite{cdfmultijets}.

This Letter presents the first results of a dedicated search for three-jet
hadronic resonances in multijet events in \Pp\Pp\ collisions. The results are
based on a data sample of proton-proton collisions at $\sqrt{s}=7$ TeV,
corresponding to an integrated luminosity of $35.1\pm1.4$~pb$^{-1}$~\cite{lumi},
collected with the Compact Muon Solenoid (CMS) detector~\cite{cmsdet} at
the Large Hadron Collider (LHC) in the running period from March
through November 2010.
Events with at least six jets, each with high transverse momentum ($p_T$),
are selected and investigated for evidence of three-jet resonances consistent
with strongly coupled supersymmetric particle decays. The event selection
criteria are optimized in the context of the gluino model mentioned above.
However, the generic features of the selection criteria provide a robust
model-independent basis that can be applied to many extensions of the SM.

The CMS detector is a multi-purpose apparatus, described in detail in
Ref.~\cite{cmsdet}.  Here, we briefly describe the subdetectors most
relevant to this analysis. The high-resolution silicon pixel and
strip tracker provides charged tracking coverage for $|\eta| < 2.4$,
where $\eta =- \ln[\tan(\theta/2)]$ is the pseudorapidity and $\theta$ is
the polar angle measured with respect to the counterclockwise
proton beam direction. Immersed in the 3.8~T magnetic field of the
superconducting solenoid, the tracker provides transverse
momentum resolution of approximately 1.5\% for charged
particles with $p_T\approx 100~\textrm{GeV}/c$. Energy deposits of the
jets are measured using electromagnetic (ECAL) and hadronic (HCAL)
calorimeters. The ECAL has a barrel part and two endcaps, is
composed of finely segmented crystals and has
an energy resolution of better than 0.5\,\% for unconverted photons with
transverse energies above 100\,GeV. The
ECAL barrel covers the pseudorapidity range $|\eta| < 1.4$ with a granularity of
$\Delta\eta \times \Delta\phi = 0.0174 \times 0.0174$, where $\phi$ is the
azimuthal angle, and the endcaps
cover $1.4 < |\eta| < 3.0$ with a granularity that decreases to
$0.05 \times 0.05$ for $|\eta| \approx 3.0$. A preshower detector
consisting of two planes of silicon sensors interleaved with a total
of three radiation lengths of lead is located in front of the ECAL endcaps.
The HCAL extends up to $|\eta| \approx 5.0$ and its central and endcap regions
consist of brass/scintillator
sampling calorimeters that cover $|\eta| < 3.0$ with a granularity of
$\Delta\eta \times \Delta\phi = 0.087 \times 0.087$ for central rapidities.
The energy of charged pions and other quasi-stable hadrons can be
measured with the calorimeters (ECAL and HCAL combined) with a
resolution of $\Delta~E/E~\approx~100~\%~\sqrt{E~\textrm{[GeV]}}\oplus~5\%$.

Events are recorded using a two-tier trigger system.
Objects satisfying the requirements at the first level (L1)
are passed to the High Level Trigger (HLT) where the total
recorded rate is limited to about ${\sim}350~$Hz. Triggers based
on the sum of all transverse energy from jets ($H_T$),
reconstructed with only calorimeter information, are
used to select recorded events. For the L1
trigger, the $H_T$ threshold is 50~GeV. The
corresponding threshold for the HLT varies between
100 and 150~GeV, depending on the run period.

The CMS particle-flow algorithm~\cite{pflow} uses calorimeter
information and combines it with reconstructed tracks to identify
individual particles such as photons, leptons, and both neutral
and charged hadrons within the jets. The energy of photons is directly
obtained from the calibrated ECAL measurement. The energy of electrons
is determined from a combination of the track momentum at the main
interaction vertex, the corresponding ECAL cluster energy, and the
energy sum of all bremsstrahlung photons attached to the track. The
energy of muons is obtained from the corresponding track momentum. The
energy of charged hadrons is determined from a combination of the track
momentum and the corresponding ECAL and HCAL energy, corrected for
zero-suppression effects, and calibrated for the nonlinear response
of the calorimeters. Finally, the energy of neutral hadrons is
obtained from the corresponding calibrated ECAL and HCAL energy.
The particle-flow objects serve as input for jet reconstruction,
performed using the anti-$k_T$ algorithm~\cite{ak5} with a distance
parameter of 0.5 in $\eta$-$\phi$ space.

Jet energy scale corrections~\cite{jec} derived from Monte
Carlo (MC) simulation are applied to account for the nonlinear
and nonuniform response of the calorimeters.
In data, a small residual correction factor is included to correct for
differences in jet response between data and simulation.
The combined corrections are on the order of 5--10\%,
and their corresponding uncertainties range from 3 to 5\%,
depending on the measured jet's
pseudorapidity and energy. Jet quality
criteria~\cite{jetid} are applied to remove
misidentified jets arising primarily from calorimeter noise.
For both data and simulated signal events, more than 99.8\%
of all selected jets satisfy these criteria.

Pair-produced gluinos are used to model the signal. Gluino production
and decay are simulated using the {\sc pythia}~\cite{pythia} MC
program (v6.420), where each gluino decays to three jets through
the $\lambda_{\textrm{uds}}$ quark RPV coupling. This coupling is
set such that the branching ratio B of the gluino to three light jets
is 100\%. The mass of the gluino is
varied between 200 and 500~GeV/$c^{2}$ in 50~GeV/$c^{2}$ steps. The
leading-order cross section from {\sc pythia} is 325~pb for
a gluino mass of 200~GeV/$c^2$, falling to $\approx$1~pb for a gluino
mass of 500~GeV/$c^2$.
For the generation of this signal all superpartners
except the gluino are taken to be decoupled~\cite{reuven}, the natural
width of the gluino resonance is taken to be much smaller than the
resolution of the detector, and no intermediate particles are produced in
the gluino decay.
The next-to-leading-order (NLO) correction factors (K factors), with
values ranging from 1.7 to 2.2, are calculated using the
{\sc prospino}~\cite{prospino} program and are applied to the leading-order
cross sections. Simulation of the CMS detector is performed
using Geant4~\cite{geant4}.

Events recorded with the $H_T$ trigger are required offline
to have a good reconstructed primary event vertex~\cite{goodPV}.
Pair-produced three-jet resonances naturally yield events
with high jet multiplicity and large transverse energy. Thus we
require events to contain at least six jets, and that the total scalar
sum of the $p_T$ of those jets is larger than 425~$\textrm{GeV}/c$.
The latter requirement ensures that the trigger is fully efficient
for these events. Jets are required to have $p_T > 45~\textrm{GeV}/c$
and $|\eta| < 3.0$, which also minimizes the effects from
multiple proton-proton interactions.

\begin{figure}
\centering
\includegraphics[width=0.95\columnwidth]{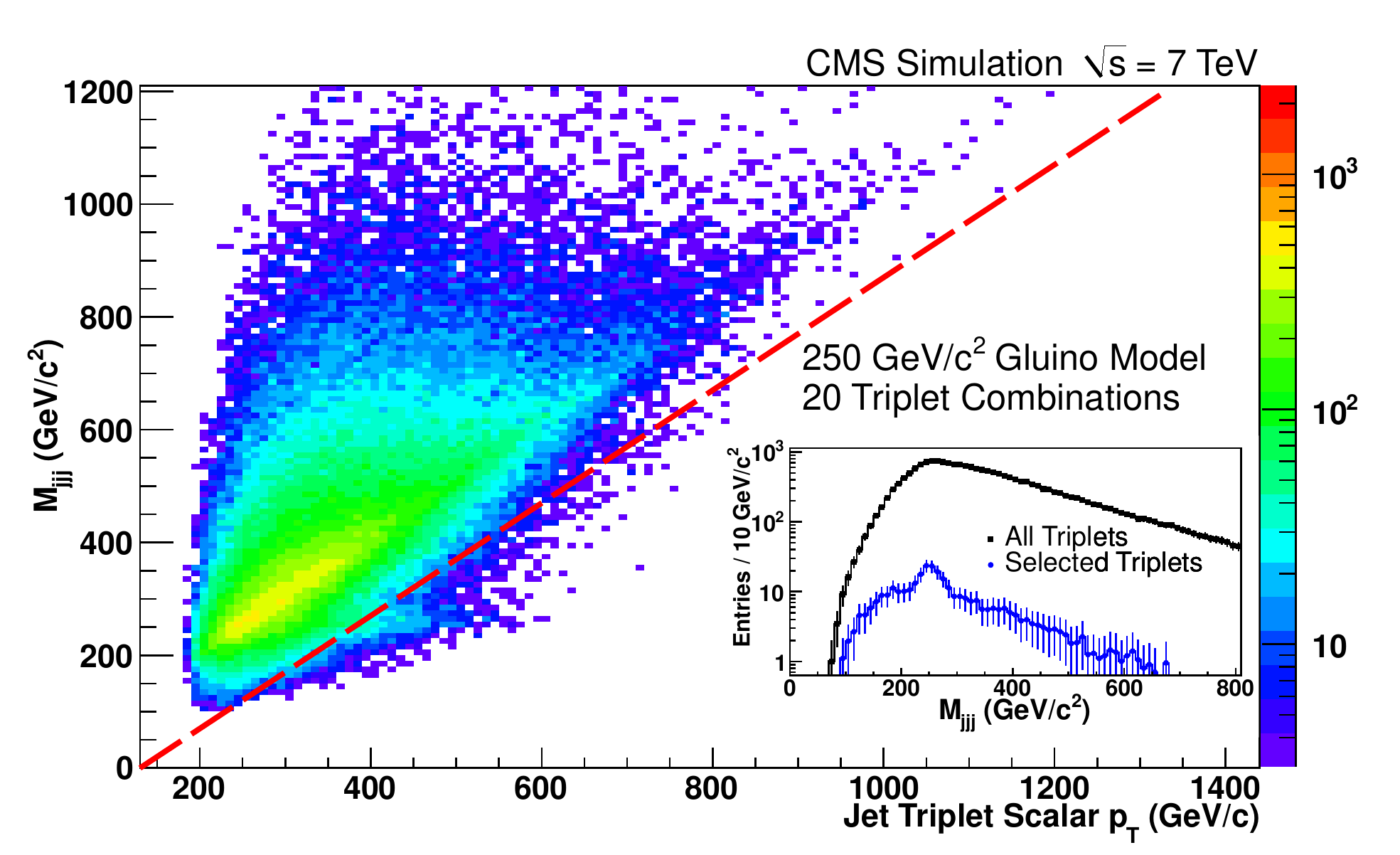}
\caption{Simulated triplet jet invariant mass $M_{jjj}$ versus
         the triplet
         scalar $p_T$ of all 20 triplets, for a
         gluino mass of 250~GeV/$c^2$. All triplets
         falling to the right of the red dashed line
         pass the requirement of Eq.~\ref{eq:diag_def}.
         In the insert, the invariant mass distribution for
         the same gluino mass is shown both before and
         after Eq.~\ref{eq:diag_def} is imposed.
}
\label{fig:2DTrips}
\end{figure}

To reconstruct the gluinos, the six highest-$p_T$ jets
are combined into all possible unique triplet combinations,
resulting in 20 combinations of jet triplets.
For signal events, each of the pair-produced
gluinos corresponds to one of these 20 jet
triplets, even in the case where
all six jets come solely from the decay of these particles,
leaving the 18 uncorrelated jet triplets as combinatorial
background. Thus, the overall background arises
not only from SM events, described by quantum chromodynamics
(QCD), but also from spurious jet triplet combinations
in signal events themselves. We impose
additional requirements on each triplet to
increase the signal sensitivity, while retaining as many
signal triplets as possible. The
invariant mass of background triplets is found to scale
with the respective scalar sum of jet $p_T$, while
for signal triplets the mass is constant.
To reduce background, we therefore require each
jet triplet to satisfy the following relation:
\begin{eqnarray}
M_{jjj} \,<\,\sum_{i=1}^{3}|p_{T}|_{i}\,-\,\Delta,
\label{eq:diag_def}
\end{eqnarray}
where $M_{jjj}$~is the triplet invariant
mass,~$\sum_{i=1}^{3}|p_{T}|_{i}$ is
the scalar sum of jet $p_T$ in the triplet (triplet scalar $p_T$),
and $\Delta$ is an offset adjusted to optimize signal
sensitivity. Figure~\ref{fig:2DTrips} shows the simulated
triplet invariant mass versus the triplet scalar
$p_T$ for a gluino mass of 250~GeV/$c^2$, and the insert
displays the invariant mass distribution before and after
the requirement. For each event, all 20 triplet combinations
are included. The value of $\Delta$ is
determined by maximizing the ratio of the number of
signal triplets to the sum of the number of signal plus
background triplets in a one standard deviation ($\sigma$)
window around the center of the gluino mass peak. A common
value of $\Delta = 130$~GeV/$c^{2}$ is taken for all gluino
masses considered, which gives an efficiency in signal
events for triplets of 1 to 5\%, and a background triplet
selection efficiency of less than 0.05\%.

Even after the final selection, background remains from
both QCD multijet events and uncorrelated triplets in gluino signal
events. The latter only contribute minimally, and the shape of
their distribution is found to be consistent with that of the
dominant background, from QCD multijet events. These QCD
multijet events arise from hard two-particle interactions combined with
initial- and final-state radiation in the form of gluon jets. Although
the cross section falls with increasing jet multiplicity ($N_\textrm{jet}$),
the underlying kinematic distributions are essentially the same among these
events. Thus, we use a rescaled mass distribution of triplets in events with
$N_\textrm{jet}= 4$, where the signal contributions are minimal,
to estimate the shape of the background.
Specifically, we select events in data with $N_\textrm{jet}= 4$
that satisfy all other selection criteria, form jet triplets, and require
each to pass Eq.~\ref{eq:diag_def}. The $M_{jjj}$ values of these
triplets are multiplied by the ratio of the average triplet scalar $p_T$
in data for events with $N_\textrm{jet}\geq 6$ to the events with $N_\textrm{jet}= 4$,
to account for expected minor kinematic differences between the two
samples. The resulting $M_{jjj}$ distribution is then fit to
an exponential function of the form: e$^{P_0 + P_1 M_{jjj} }$, where
$P_0$ and $P_1$~are free parameters. The slope $P_1$ of the exponential function
in the
$N_\textrm{jet}\geq 6$ sample is constrained to be equal to that found for
the scaled $N_\textrm{jet}= 4$ fit within its uncertainties. This is verified
in QCD simulation, and as a cross-check in data, we apply this
procedure to predict the shape of the $M_{jjj}$ distribution for an
$N_\textrm{jet}= 5$ sample, where the QCD multijet background is also expected to
dominate, and find good agreement.
To verify that the choice of the background model does not bias the derived limit,
the exponential function is tested on an $N_\textrm{jet}\geq 6$ sample,
defined by the standard selection criteria without the requirement of
Eq.~\ref{eq:diag_def} imposed. The parameterization is found to be in agreement
with the data in the fitted region, with the slope of the fit consistent with
those of the $N_\textrm{jet}= 4$ and $N_\textrm{jet}= 5$ samples.

To estimate the number of signal events expected after
all selection criteria are applied, the sum of a
Gaussian function that represents the signal and the
exponential function that models the background is
fitted to the simulated $M_{jjj}$ distribution for
each gluino mass. The
fit is performed in the range  $170 < M_{jjj} < 800$~GeV/$c^2$.
The width of the Gaussian function modeling the signal
varies according to the detector resolution, and
gluinos of mass from 200 to 500~GeV/$c^2$ correspond
to widths from 10 to 25~GeV/$c^{2}$.
The integral of the Gaussian component provides the estimate
for the expected number of signal triplets produced, and
the value of this integral, divided by the number of signal
events generated, determines the signal acceptance for each
gluino mass.
The signal acceptance is parameterized using a second-order
polynomial as a function of gluino mass, and
the acceptance ranges from 0.4 to 5\% as the gluino mass
increases from 200 to 500~GeV/$c^2$.

The systematic uncertainty on the signal acceptance
is evaluated in the following way.
An uncertainty related to the jet energy scale~\cite{jec} is
evaluated by varying the jet energy scale correction within its
uncertainties,
then recalculating the acceptance for different gluino mass values.
The largest difference with respect to the nominal
acceptance is taken as the systematic uncertainty and ranges from
7 to 16\%. The amount of initial- and final-state radiation is
increased and decreased according to a standard
prescription~\cite{topmass}, and the associated uncertainty is
quantified in a similar manner to that described for the jet energy
scale uncertainties.
The difference of 2 to 4\% with respect to the nominal acceptance
is taken as the systematic uncertainty. To determine the effects of
additional proton-proton interactions on the signal acceptance, signal
samples are generated with the number of interactions per bunch
crossing in the simulation set to the average of their distribution
in the data. Applying the acceptance calculation on this sample
leads to differences of 1 to 6\%, which are taken as uncertainties.
These contributions, combined with those from the luminosity
measurement (4\%) and choice of parton distribution function set
(4\%), give a total systematic uncertainty on the signal yield
between 10 and 19\%, depending on the value of the gluino mass.
Other effects, such as additional background parameterizations
and variations of the fit range, were also tested and found to
be negligible.

\begin{figure}
\centering
\includegraphics[width=0.95\columnwidth]{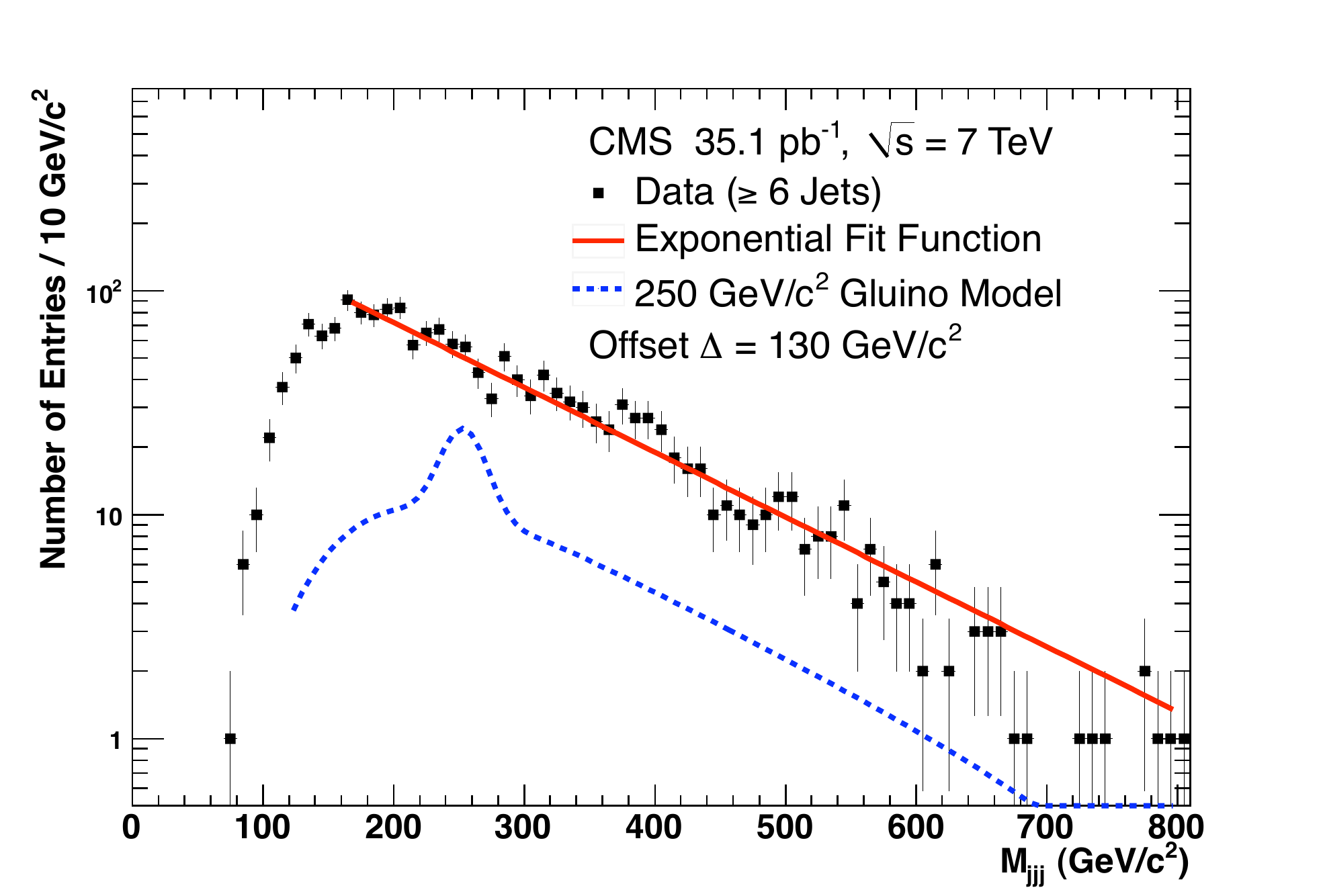}
\caption{Three-jet invariant mass distribution of
         triplets passing all selection criteria
         for the $N_\textrm{jet}\ge 6$ data sample.
         An exponential
         function representing the background shape,
         constrained from the $N_\textrm{jet}= 4$
         distribution, and the expectation for the
         250~GeV/$c^{2}$ gluino signal are also shown.
}
\label{fig:DataFitB}
\end{figure}

Figure~\ref{fig:DataFitB} shows the three-jet invariant mass distribution for
the $N_\textrm{jet}\ge 6$ sample with all selection criteria applied, and
the exponential fit superimposed. The simulated signal distribution for a
gluino mass of 250~GeV/$c^2$, normalized to the integrated luminosity of the
data sample, is also shown. Because agreement is observed between the data
and expected QCD background, a limit-setting procedure is performed.

Upper limits are placed on the cross section $\sigma_S$ for the production
of three-jet resonances in the $N_\textrm{jet}\ge~6$ sample using a Bayesian
approach. The background model parameters and their corresponding
uncertainties are taken from the fit of the exponential function to the
$N_\textrm{jet}\geq~6$ distribution, constrained by the $N_\textrm{jet}= 4$
sample, with all selection criteria applied.
The uncertainties on the two parameters that describe the background
shape, namely the exponential slope and normalization, are included as
Gaussian priors. The central value is set to the best fit value and the
width to one standard deviation. The range is truncated at $\pm 3 \sigma$.
In addition to the background parameters, priors are included for the
acceptance and integrated luminosity.

The integrated luminosity, acceptance, signal width and the two parameters
of the exponential background distribution are all treated as nuisance
parameters. The likelihood is combined with the prior and nuisance parameters,
and then marginalized to give the posterior density for $\sigma_S$.
Integrating the posterior density to 0.95 of the total gives the 95\%
confidence level (CL) limit for $\sigma_S$. Marginalization and integration
of the posterior density are performed with a Markov chain MC integration
technique using RooStats~\cite{roostats}.

To determine the expected limits, a large set of pseudo-experiments (PEs)
is generated using the background-only model. For every PE, each of the
two parameters associated with the exponential is varied by generating a
random number distributed according to a Gaussian probability distribution
function centered at the central value, with a width corresponding to the
associated uncertainty. The total number of events in a given PE is
extracted according to the Poisson distribution with mean value equal to
the number of events predicted by the exponential function
in the fitted range. The same upper limit calculation
performed on data is repeated for each PE at each mass, and the
median of the upper limit distribution for all PEs is the expected
limit.

\begin{figure}
\centering
\includegraphics[width=0.95\columnwidth]{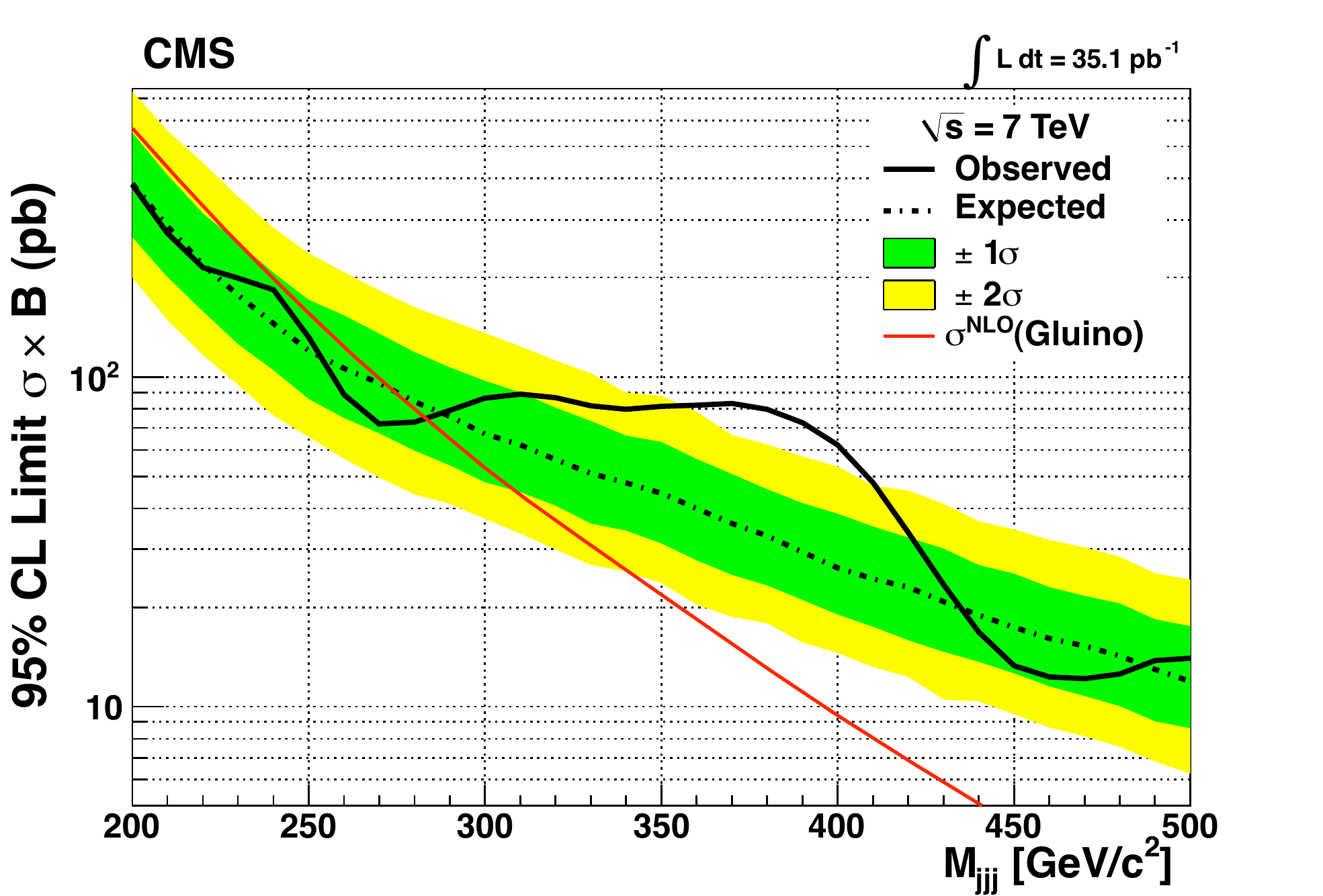}
\caption{Observed and expected 95\% CL upper limits on the
cross section for gluino pair production through RPV decays, where the
branching ratio of the gluino to three jets is 100\%. Also shown are the
$\pm1\sigma$ and $\pm2\sigma$ bands on the expected limit, as well
as the theoretical NLO cross section for gluino production.}
\label{fig:Limits}
\end{figure}

The observed and expected 95\% CL upper limits on the gluino pair
production cross section times branching ratio as a function of gluino
mass are presented in Fig.~\ref{fig:Limits} and Table~\ref{tbl:Limits}.
The corresponding 95\% CL lower limit on the gluino mass is set by finding
the mass value at which the 95\% CL limit line crosses that of the NLO
gluino cross section. We thus exclude at the 95\% CL gluino masses in the
range 200 to 280~GeV/$c^2$, with an expected lower limit of 270~GeV/$c^2$.
The most significant excess occurs for a mass around 390~GeV/$c^{2}$,
corresponding to a significance of 1.9 standard deviations, when the
so-called look-elsewhere effect~\cite{lookelse} is taken into account.

\begin{table*}
\centering
\caption{Observed and expected 95\% CL upper limits on the
cross section times branching ratio for the pair production
of gluinos with masses ($M_{jjj}$) ranging from 200 to
500~GeV/$c^{2}$.}
\begin{tabular}{ccc|ccc}
\hline
\hline
$M_{jjj}$ (GeV/$c^2$) & Observed (pb) & Expected (pb) & $M_{jjj}$ (GeV/$c^2$) & Observed (pb) & Expected (pb) \\
\hline
200   & 383   & 387 & 360   & 82  & 40 \\
210   & 273   & 287 & 370   & 83  & 36 \\
220   & 214   & 219 & 380   & 80  & 33 \\
230   & 200   & 178 & 390   & 73  & 29 \\
240   & 184   & 146 & 400   & 62  & 26 \\
250   & 132   & 120 & 410   & 48  & 24 \\
260   & 88    & 106 & 420   & 34  & 23 \\
270   & 72    & 96  & 430   & 24  & 21 \\
280   & 73    & 84  & 440   & 17  & 19 \\
290   & 79    & 76  & 450   & 13  & 17 \\
300   & 86    & 67  & 460   & 12  & 16 \\
310   & 89    & 62  & 470   & 12  & 15 \\
320   & 87    & 56  & 480   & 13  & 14 \\
330   & 82    & 51  & 490   & 14  & 13 \\
340   & 80    & 48  & 500   & 14  & 12 \\
350   & 82    & 45  &       &       &  \\
\hline
\end{tabular}
\label{tbl:Limits}
\end{table*}

In summary, a search for three-jet hadronic resonance production in \Pp\Pp\ collisions at a center-of-mass energy of 7 TeV has been conducted by the
CMS Collaboration, using a data sample corresponding to 35~pb$^{-1}$. Events
having the properties of high jet multiplicity and large scalar sum of jet
$p_{T}$, which are expected signatures of high mass hadronic resonances,
are analyzed for the presence of signal events with a model-independent
approach. The number of expected SM background events is found to be in
good agreement with the observed events. The production of gluinos decaying
through the $\lambda_{\textrm{uds}}$ RPV coupling is excluded for masses
between 200 and 280~GeV/$c^{2}$ at 95\% CL. These are complementary
to recent results from the Tevatron, which rule out gluino masses below
144~GeV/$c^2$~\cite{cdfmultijets}, and these limits are the first
from a dedicated search of this kind in \Pp\Pp\ collisions.

The authors would like to thank Michael Park and Yue Zhao for providing
theoretical calculations. We wish to congratulate our colleagues in the CERN
accelerator departments for the excellent performance of the LHC machine.
We thank the technical
and administrative staff at CERN and other CMS institutes, and acknowledge
support from: FMSR (Austria); FNRS and FWO (Belgium); CNPq, CAPES, FAPERJ,
and FAPESP (Brazil); MES (Bulgaria); CERN; CAS, MoST, and NSFC (China);
COLCIENCIAS (Colombia); MSES (Croatia); RPF (Cyprus); Academy of Sciences
and NICPB (Estonia); Academy of Finland, ME, and HIP (Finland); CEA and
CNRS/IN2P3 (France); BMBF, DFG, and HGF (Germany); GSRT (Greece); OTKA and
NKTH (Hungary); DAE and DST (India); IPM (Iran); SFI (Ireland); INFN
(Italy); NRF and WCU (Korea); LAS (Lithuania); CINVESTAV, CONACYT, SEP,
and UASLP-FAI (Mexico); PAEC (Pakistan); SCSR (Poland); FCT (Portugal);
JINR (Armenia, Belarus, Georgia, Ukraine, Uzbekistan); MST and MAE
(Russia); MSTD (Serbia); MICINN and CPAN (Spain); Swiss Funding Agencies
(Switzerland); NSC (Taipei); TUBITAK and TAEK (Turkey); STFC
(United Kingdom); DOE and NSF (USA).

\bibliography{auto_generated}   % will be created by the tdr script.
\cleardoublepage \appendix\section{The CMS Collaboration \label{app:collab}}\begin{sloppypar}\hyphenpenalty=5000\widowpenalty=500\clubpenalty=5000\textbf{Yerevan Physics Institute,  Yerevan,  Armenia}\\*[0pt]
S.~Chatrchyan, V.~Khachatryan, A.M.~Sirunyan, A.~Tumasyan
\vskip\cmsinstskip
\textbf{Institut f\"{u}r Hochenergiephysik der OeAW,  Wien,  Austria}\\*[0pt]
W.~Adam, T.~Bergauer, M.~Dragicevic, J.~Er\"{o}, C.~Fabjan, M.~Friedl, R.~Fr\"{u}hwirth, V.M.~Ghete, J.~Hammer\cmsAuthorMark{1}, S.~H\"{a}nsel, M.~Hoch, N.~H\"{o}rmann, J.~Hrubec, M.~Jeitler, W.~Kiesenhofer, M.~Krammer, D.~Liko, I.~Mikulec, M.~Pernicka, B.~Rahbaran, H.~Rohringer, R.~Sch\"{o}fbeck, J.~Strauss, A.~Taurok, F.~Teischinger, P.~Wagner, W.~Waltenberger, G.~Walzel, E.~Widl, C.-E.~Wulz
% ...
\vskip\cmsinstskip
% ...
1:~~Also at CERN, European Organization for Nuclear Research, Geneva, Switzerland\\
% ...

\end{sloppypar}
\end{document}